\documentclass[sigconf]{acmart}
\AtBeginDocument{%
  \providecommand\BibTeX{{%
    \normalfont B\kern-0.5em{\scshape i\kern-0.25em b}\kern-0.8em\TeX}}}
\usepackage{float} 
\copyrightyear{2023}
\acmYear{2023}
\setcopyright{acmlicensed}\acmConference[BCB '23]{14th ACM International
Conference on Bioinformatics, Computational Biology and Health
Informatics}{September 3--6, 2023}{Houston, TX, USA}
\acmBooktitle{14th ACM International Conference on Bioinformatics,
Computational Biology and Health Informatics (BCB '23), September 3--6,
2023, Houston, TX, USA}
\acmPrice{15.00}
\acmDOI{10.1145/3584371.3612951}
\acmISBN{979-8-4007-0126-9/23/09}

\begin{document}
\title{RoseNet: Predicting Energy Metrics of Double InDel Mutants Using Deep Learning}

 \author{Sarah Coffland}
 \email{cofflas@wwu.edu}
 \affiliation{%
   \institution{Western Washington University}
   \streetaddress{516 High St}
   \city{Bellingham}
   \state{Washington}
   \country{USA}
   \postcode{98225}
 }

 \author{Katie Christensen}
 \email{chris90@wwu.edu}
 \affiliation{%
   \institution{Western Washington University}
   \streetaddress{516 High St}
   \city{Bellingham}
   \state{Washington}
   \country{USA}
   \postcode{98225}}

 \author{Filip Jagodzinski}
 \email{filip.jagodzinski@wwu.edu}
 \affiliation{%
   \institution{Western Washington University}
   \streetaddress{516 High St}
   \city{Bellingham}
   \state{Washington}
   \country{USA}
   \postcode{98225}}

 \author{Brian Hutchinson}
 \email{Brian.Hutchinson@wwu.edu}
 \affiliation{%
   \institution{Western Washington University}
   \streetaddress{516 High St}
   \city{Bellingham}
   \state{Washington}
   \country{USA}
   \postcode{98225}}

\begin{abstract}
    An amino acid insertion or deletion, or InDel, can have profound and varying functional impacts on a protein's structure. InDel mutations in the transmembrane conductor regulator protein for example give rise to cystic fibrosis. Unfortunately performing InDel mutations on physical proteins and studying their effects is a time prohibitive process. Consequently, modeling InDels computationally can supplement and inform wet lab experiments. In this work, we make use of our data sets of exhaustive double InDel mutations for three proteins which we computationally generated using a robotics inspired inverse kinematics approach available in Rosetta. We develop and train a neural network, RoseNet, on several structural and energetic metrics output by Rosetta during the mutant generation process. We explore and present how RoseNet is able to emulate the exhaustive data set using deep learning methods, and show to what extent it can predict Rosetta metrics for unseen mutant sequences with two InDels. RoseNet achieves a Pearson correlation coefficient median accuracy of 0.775 over all Rosetta scores for the largest protein. Furthermore, a sensitivity analysis is performed to determine the necessary quantity of data required to accurately emulate the structural scores for computationally generated mutants. We show that the model can be trained on minimal data ($<50\%$) and still retain a high level of accuracy.

\end{abstract}

\maketitle

\section{Introduction}
Wet lab studies, while extremely useful, are not feasible for conducting comprehensive experiments of all possible mutants with one, let alone two InDel mutations. Even a modest-sized protein with 100 residues can yield $101 \times 20 = 2,020$ unique mutants with a single insertion mutation, and ${n=101 \choose 2} \times 20^2 = 2,020,000$ unique mutants with two insertion mutations. Engineering such large counts of physical proteins and then performing experiments on them is infeasible. Consequently research turns to computational approaches for modeling and analyzing proteins and protein mutants. Scores generated by the Rosetta software evaluating the energy and stability of different protein conformations are now an instrumental part of structural bioinformatics research~\cite{alford2017rosetta}.

We have expanded a previous compute pipeline~\cite{turcan2022exhaustive} that relies on Rosetta~\cite{leaver2011rosetta3} to generate \textit{in silico} all mutants with two InDels for three protein PDB structure files.

Computationally generating a protein mutant with two InDels can require upwards of several minutes of compute time. Generating an exhaustive set of upwards of 2,000,000 mutants can thus require upwards of 3000 days if tasked on a single CPU, or several weeks if performed via a distributed fashion across hundreds of CPUs, as we have done. Thus even using a computational approach does not permit real-time experiments that seek to elucidate trends of the effects of InDels for all possible mutants.

In this work, we train a deep learning model called RoseNet, modeled after the ResNet architecture for image recognition~\cite{he2016deep}. We replace the convolutional layers in ResNet with fully-connected layers, and predict the Rosetta scores for all InDel mutants without requiring the comprehensive list of mutants PDB structures. In addition, a sensitivity analysis is performed to ascertain the amount of data that is required to effectively emulate the complete set of structural scores for the mutants generated by the Rosetta software. 

\section{Related Work}
Wet-lab and computational efforts studying the effects of substitution mutations have been the focus of research efforts dating back to the 1970s~\cite{drake1976biochemistry,masso2008accurate,moreira2007computational}. Experimental techniques including alanine scanning shotgun approaches which permit high-throughput analyses~\cite{morrison2001combinatorial} of large data sets of mutant variants have in part enabled those efforts. Research on simulating and predicting the effects of InDels also is available, albeit not as plentiful~\cite{savino2022insertions,choi2012predicting}.

Recent approaches related to this research include investigations into the impacts of InDels and particularly their influence on disease phenotypes, through computational modeling which involves comparing InDel-induced structural effects with those from amino acid substitutions~\cite{jilani2022elucidating}. In the most recent related work, InDels were computationally generated and the single InDel mutations on protein structures were assessed quantitatively~\cite{turcan2022exhaustive}.  Nonetheless there is still significant progress needed in computational approaches so that they can be adopted widely, as shown by Miosage \textit{et al.} in~\cite{miosage2015}, who revealed discrepancies between observed and predicted impacts made by PolyPhen2, SIFT, and other tools.

\section{Methods}
The proteins studied here were the 99-residue aspartyl protease from the HIV-1 isolate BRU (PDB 1hhp), the 67-residue \textit{Bacillus Subtilis} Major Cold Shock Protein (PDB 1csp), and the 46-residue Cambrin protein (PDB 1crn). Each of the three PDB structure files was input to our compute pipeline~\cite{turcan2022exhaustive}, which invokes Rosetta, to generate a comprehensive set of mutant structure files with two InDels across all possible combinations of the twenty naturally occurring amino acids across the various insertion positions along the amino acid chain. 

For this work we rely on the exhaustive data sets of pairwise InDel mutants for several proteins. To generate the data set, we extended our previous compute pipeline that relies on the inverse kinematics inspired methods in Rosetta to make two insertions, and which then closes the loops at the insertion points to yield a protein mutant~\cite{das2008macromolecular,turcan2022exhaustive}. During the mutant generation process a variety of energetic and related metrics are output by Rosetta (Table~\ref{tab:RosettaScores}). These metrics characterize the mutant generation process, and offer various quantitative measures of a generated mutant structure with two InDels.

Each residue in every mutant contains all twenty Rosetta scores. In order to make use of the scores to characterize the entire protein structure, each of the Rosetta scores were summed for all residues, to yield a composite Rosetta metric. It was noticed that when the two InDel locations were within 12 residues of each other in the protein's sequence, the Rosetta software often was unable to generate a mutant within the maximum iterative time step stipulated via the input parameters to the software. The reasons for a mutant not being generated within a specified time frame was most often due to steric clashes imposed by the inserted residues, which prevented the inverse kinematic chain model from closing the gap where the insertions were made. Due to this, we did not make use of the mutant data of InDels within 12 residues of each other. In addition, two of the Rosetta scores (dslf\_fa13 and yhh\_planarity) were omitted from training and evaluation due to their lack of interest as both metrics were constant over all mutants.

\begin{table}
    \centering
    \caption{Energetic and related metrics output by Rosetta. e=energy, res=residue.}
    \begin{tabular}{lp{5.5cm}}
         \toprule
            Rosetta score/metric & Meaning \\
            \midrule
            fa\_atr   & Lennard-Jones attractive force between atoms in different res \\
             fa\_rep   & Lennard-Jones repulsive force between atoms in different res \\
            fa\_sol   & Lazaridis-Karplus solvation e \\
            fa\_intra\_sol\_xover4   & Intra-residue Lazaridis-Karplus solvation e \\
            lk\_ball\_wtd   & Asymmetric solvation e \\
            fa\_intra\_rep   & Lennard-Jones repulsive between atoms in the same res \\
            fa\_elec   & Coulombic electrostatic potential with distance-dependent dielectric  \\
            pro\_close   & Proline ring closure e and e of $\psi$ angle of preceding residue \\
            hbond\_sr\_bb   & Backbone-backbone hbonds close in primary sequence \\
            hbond\_lr\_bb   & Backbone-backbone hbonds distant in primary sequence \\
            hbond\_bb\_sc   & Sidechain-backbone hydrogen bond e \\
            hbond\_sc   & Sidechain-sidechain hydrogen bond e \\
            dslf\_fa13   & Disulfide geometry potential \\
            rama\_prepro   & Ramachandran preferences \\
            omega   & 	Omega dihedral in the backbone\\
            p\_aa\_pp   & Probability of amino acid, given torsion values for $\phi$ and $\psi$ \\
            fa\_dun   & Internal energy of sidechain rotamers \\
            yhh\_planarity   & Torsional potential to keep tyrosine hydroxyl in plane of aromatic ring \\
            dslf\_fa13   & Disulfide geometry potential \\
            ref   & Reference e for each amino acid\\
            total   &  The total weighted score for the structure \\
  \bottomrule
    \end{tabular}
    \label{tab:RosettaScores}
\end{table}

\begin{figure}
\includegraphics[width=0.50\textwidth]{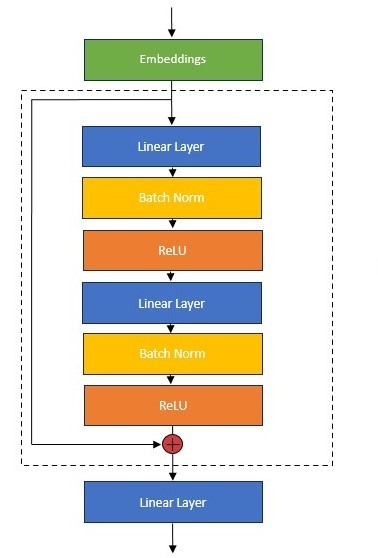}
\caption{RoseNet Architecture diagram detailing the contents of a single RoseNet block.}
\label{fig:rosenetblock}
\end{figure}

Our model (Figure~\ref{fig:rosenetblock}) consists of a deep neural network (DNN) loosely modeled after the ResNet architecture introduced by He \textit{et al.}~\cite{he2016deep}. Instead of the convolutional layers and max and average pooling layers typically used in computer vision tasks, only fully-connected layers, batch normalization and ReLU activation functions were used (as this task lacks the spatial nature that convolutional layers are design for). During training the model takes in four inputs: the first and second amino acid identities, and the insertion positions of each of those amino acids. These four inputs are turned into low dimensional vectors via embedding layers prior to being fed into the remainder of the model. The model and training was implemented using the PyTorch library~\cite{paszke2019pytorch}, and the criterion for the loss function is Huber loss ~\cite{huberloss}.

To train and evaluate our model, the data was split as follows into training, validation, and three test sets. 
For a protein composed of $T$ residues, we begin with a tensor ${\bf X}$, where $\bf{X}_{i,j,k}$th entry contains the $k$th Rosetta score when the $i$th amino-acid pair is inserted at position-pair $j$. Note that $i \in I_1 = \{1,2,\dots,400\}$ because there are $20\cdot 20=400$ possible amino acid pairs; $j \in I_2 = \{1,2,\dots,(T-12)(T-11)/2\}$ (enumerating the valid pairs of insertion locations for double insertions where the second insertion is at least 12 greater than the first), and $k \in I_3 = \{1,2,\dots,18\}$, after excluding the two constant Rosetta scores.
To assess generalization to previously unseen amino acid combinations, we define ``Test3'' to be $\{ {\bf X}_{i,j,k} : i \in R_1, j \in I_2, k \in I_3 \}$, where $R_1$ is a random 5\% subset of $I_1$.
To assess generalization to previously unseen insertion location pairs, we define ``Test2'' to be $\{ {\bf X}_{i,j,k} : i \in I_1, j \in R_2, k \in I_3 \}$, where $R_2$ is a random 5\% subset of $I_2$.
Note that while Test2 and Test3 each account for $\sim 5$\% of the total data, they are partially overlapping by definition. All of the remaining elements of the tensor, excluding those in Test2 and Test3, are randomly partitioned at a 80\%-10\%-10\% ratio into Training, Validation and Test1, respectively. This final test set, Test1, is used to assess generalization to double insertions where both the amino acid pair and location pair have been previously seen, but not together. The training set is used to train the neural network parameters, while the validation set is used to tune model and training hyperparameters. Among the training hyperparameters we tuned was the learning rate, and the optimizer chosen for best performance was NAdam. 

For each protein considered in this study, we also performed a sensitivity analysis to determine the effect of training set size on Rosetta score prediction. To perform this study, we consider a series of proper subsets of the training data at different fractions of the total size, and evaluate performance at each.

\section{Results}
While the generation of the Rosetta scores from which the model was trained took several weeks of continuous computation for each protein employing a distributed approach where Rosetta jobs were distributed across multiple CPUs, once these results were properly stored the model trained very quickly. 
The three proteins fully trained in less than 10 epochs whether 1, 5 or 10 RoseNet blocks were used (Table~\ref{tab:times}). Our experimental results demonstrate that RoseNet achieved comparable performance on the test set and the training set, indicating no significant overfitting or underfitting concerns.

\begin{table}
  \caption{Time per epoch of training (minutes) for 1, 5 and 10 RoseNet blocks respectively.}
  \label{tab:times}
  \begin{tabular}{cccl}
    \toprule
    Protein&1 block&5 blocks&10 blocks\\
    \midrule
        1hhp   & 1:25  & 2:48  & 4:45 \\
        1csp   & 0:37  & 1:10  & 1:57 \\
        1crn   & 0:14  & 0:27  & 0:46 \\
  \bottomrule
\end{tabular}
\end{table}

\subsection{Results - 1hhp}

1hhp, which is the 99-residue aspartyl protease from HIV-1, is the largest PDB structure that we made use of. On a single GPU, 1hhp takes 1 minute, 25 seconds per epoch to train, and trains fully in 7 epochs.
The best performance on our validation set was a model that contained only one RoseNet block.

The test results are shown in Table~\ref{tab:1hhptest}. Recall that the first test set used to evaluate the performance of RoseNet was comprised of data points where the amino-acid pair and insertion-location pairs had both been seen individually, but not together. On Test1, the scores range from 0.431 for fa\_intra\_rep to 1.000 for the referential energy. The median score is 0.775 and the mean of the scores is 0.766. Test2 assesses generalization to previously unseen insertion location pairs. On this second test set, the scores range from a minimum of 0.407 to a maximum of 1.000. The median score is 0.764 and the mean of all of the scores is 0.746. It is noteworthy that there is minimal loss of performance when encountering previously unseen insertion location pairs.
The third test set assesses generalization to new amino-acid combinations. On Test3, the scores range from a minimum of 0.437 to a maximum of 1.000. The median score is 0.768 and the mean of all of the scores is 0.765. Generalizing to new amino-acid pairs seems to be roughly as difficult as generalizing to insert-location pairs.

\begin{table}
  \caption{Tests 1, 2 and 3 Pearson correlation coefficients for all predicted scores for 1hhp.}
  \label{tab:1hhptest}
  \begin{tabular}{llll}
    \toprule
    Rosetta Score/Metric & Test1 & Test2 & Test3 \\
    \midrule
    fa\_atr & 0.722 & 0.689 & 0.713 \\
    fa\_rep & 0.689 & 0.657 & 0.695 \\
    fa\_sol & 0.780 & 0.763 & 0.776 \\
    fa\_intra\_rep & 0.431 & 0.407 & 0.437 \\
    fa\_intra\_sol & 0.770 & 0.777 & 0.761 \\
    lk\_ball\_wtd & 0.755 & 0.738 & 0.754 \\
    fa\_elec & 0.823 & 0.767 & 0.823 \\
    pro\_close & 0.580 & 0.571 & 0.575 \\
    hbond\_sr\_bb & 0.905 & 0.867 & 0.903 \\
    hbond\_lr\_bb & 0.873 & 0.830 & 0.873 \\
    hbond\_bb\_sc & 0.808 & 0.817 & 0.812 \\
    hbond\_sc & 0.886 & 0.865 & 0.889 \\
    omega & 0.800 & 0.765 & 0.796 \\
    fa\_dun & 0.870 & 0.858 & 0.864 \\
    p\_aa\_pp & 0.744 & 0.739 & 0.742 \\
    ref & 1.000 & 1.000 & 1.000 \\
    rama\_prepro & 0.659 & 0.643 & 0.649 \\
    total & 0.700 & 0.667 & 0.706 \\
    \bottomrule
  \end{tabular}
\end{table}

\begin{table}
  \caption{Sensitivity analysis showing the $R^2$ metrics for 1hhp for Tests 1, 2, and 3.}
  \label{tab:1hhpr2}
  \begin{tabular}{llll}
    \toprule
    Percent of Data & Test1 & Test2 & Test3 \\
    \midrule
    100 & 0.602 & 0.551 & 0.584 \\
    90 & 0.600 & 0.547 & 0.582 \\
    80 & 0.600 & 0.549 & 0.582 \\
    70 & 0.601 & 0.549 & 0.582 \\
    60 & 0.599 & 0.549 & 0.582 \\
    50 & 0.598 & 0.548 & 0.58 \\
    40 & 0.593 & 0.545 & 0.574 \\
    30 & 0.592 & 0.538 & 0.574 \\
    20 & 0.586 & 0.533 & 0.568 \\
    10 & 0.565 & 0.512 & 0.544 \\
    5 & 0.543 & 0.488 & 0.521 \\
    1 & 0.426 & 0.368 & 0.400 \\
    0.500 & 0.474 & 0.420 & 0.454 \\
    0.100 & 0.397 & 0.337 & 0.373 \\
    \bottomrule
  \end{tabular}
\end{table}

\begin{figure*}
\caption{One-to-one plots showing the performance on Test1 for  1hhp. The x-axis denotes the true values and the y-axis denotes the predictions made by our RoseNet model} \label{fig:1hhp_one2one}
\includegraphics[width=0.24\textwidth]{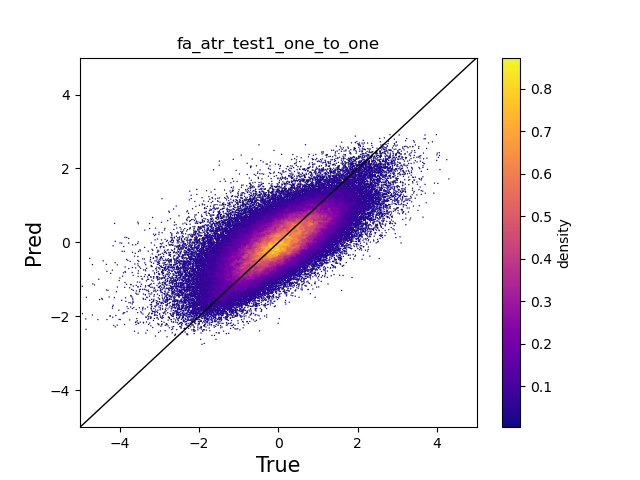}
\includegraphics[width=0.24\textwidth]{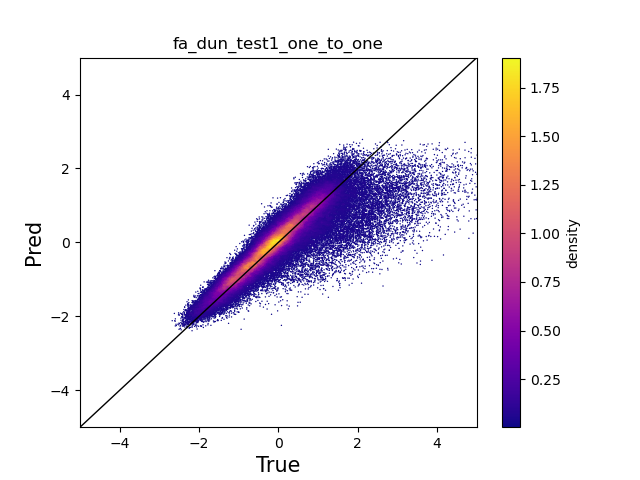}
\includegraphics[width=0.24\textwidth]{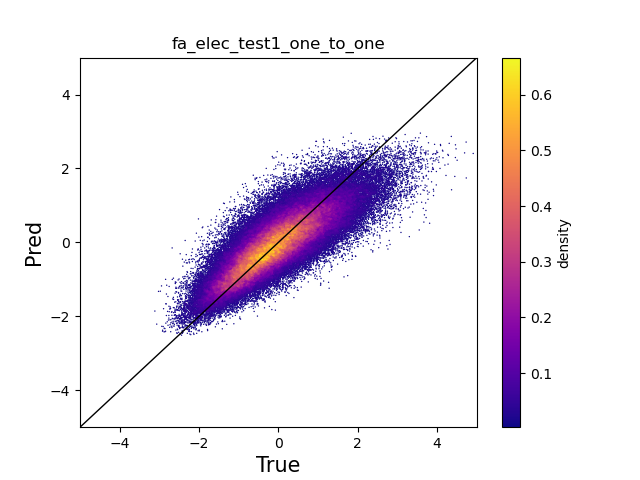}
\includegraphics[width=0.24\textwidth]{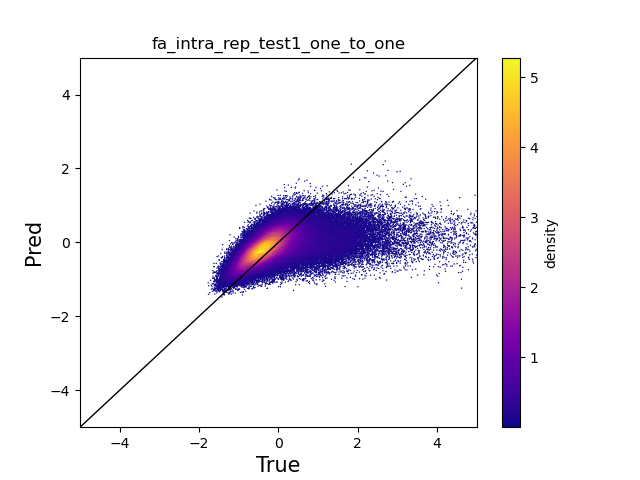}
\includegraphics[width=0.24\textwidth]{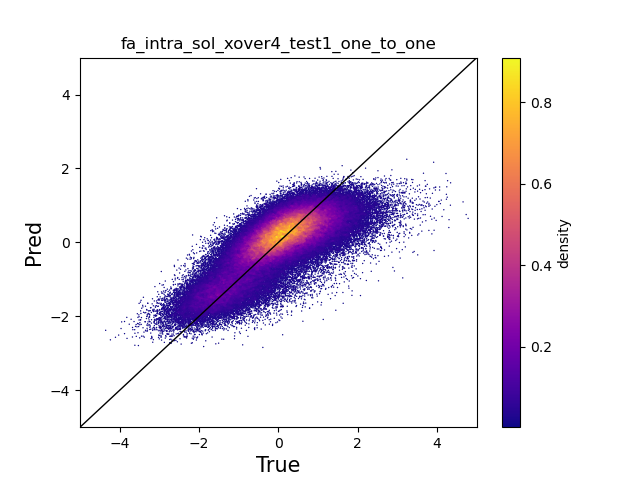}
\includegraphics[width=0.24\textwidth]{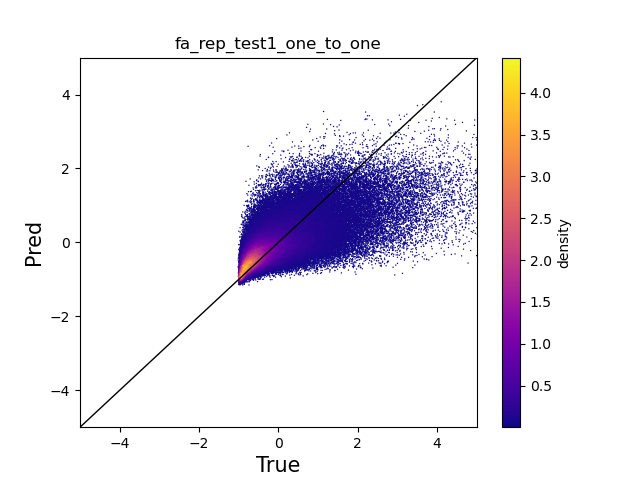}
\includegraphics[width=0.24\textwidth]{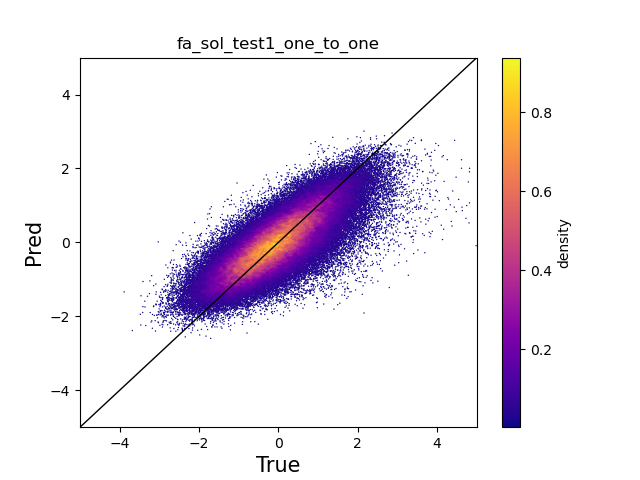}
\includegraphics[width=0.24\textwidth]{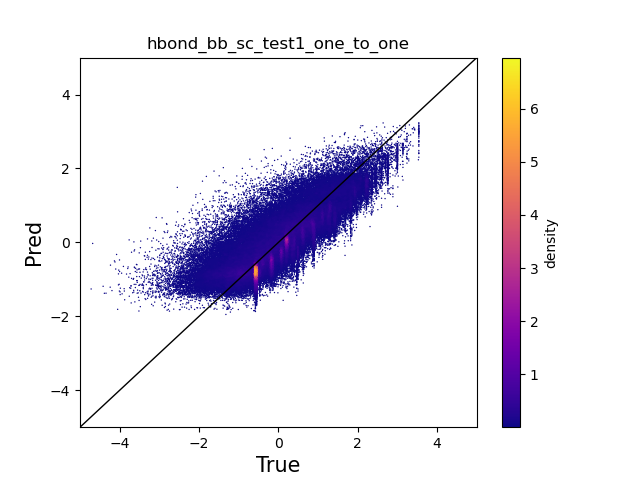}
\includegraphics[width=0.24\textwidth]{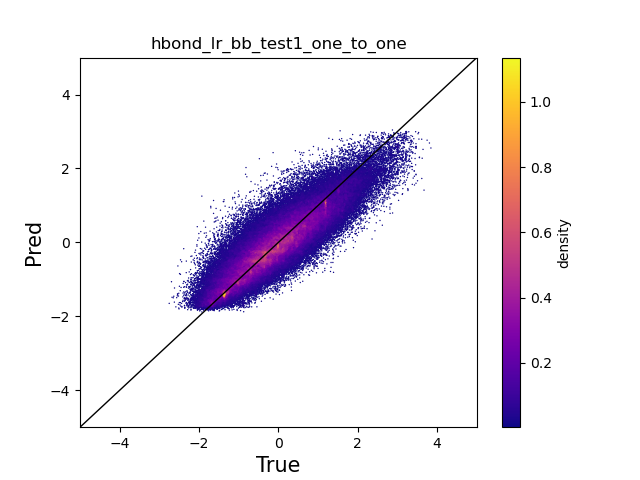}
\includegraphics[width=0.24\textwidth]{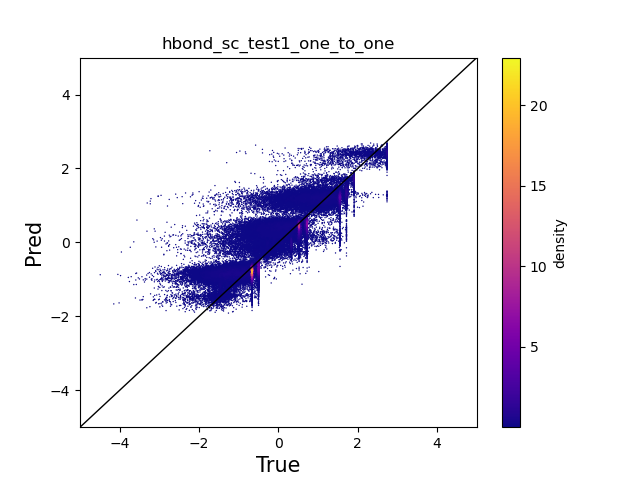}
\includegraphics[width=0.24\textwidth]{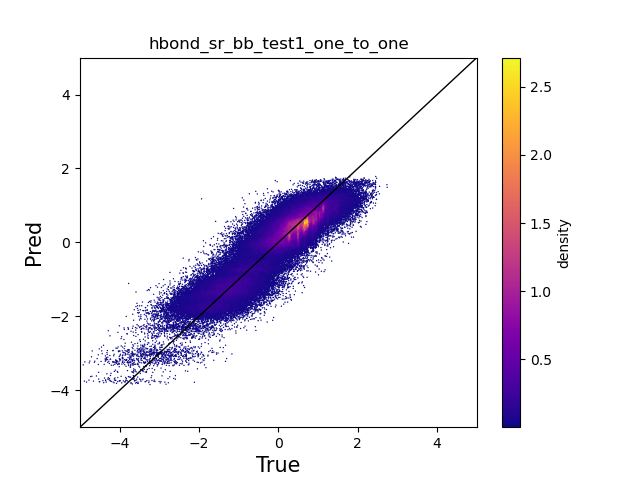}
\includegraphics[width=0.24\textwidth]{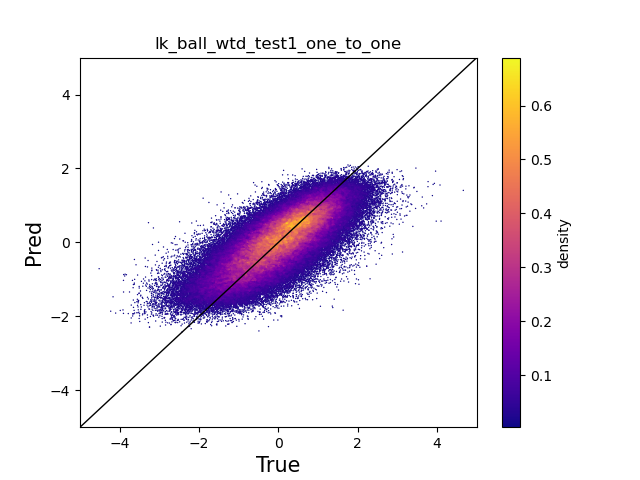}
\includegraphics[width=0.24\textwidth]{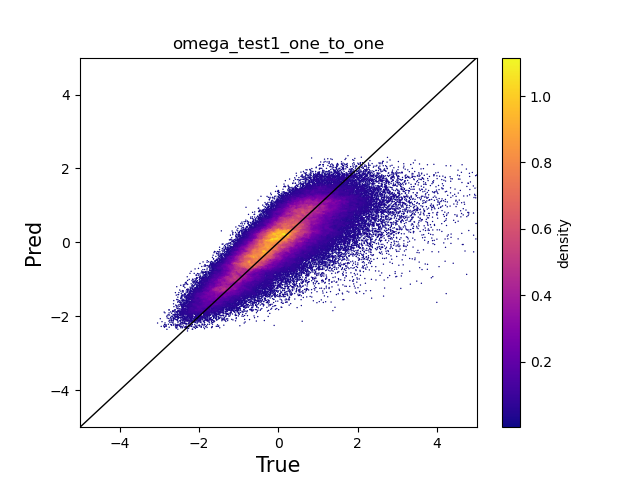}
\includegraphics[width=0.24\textwidth]{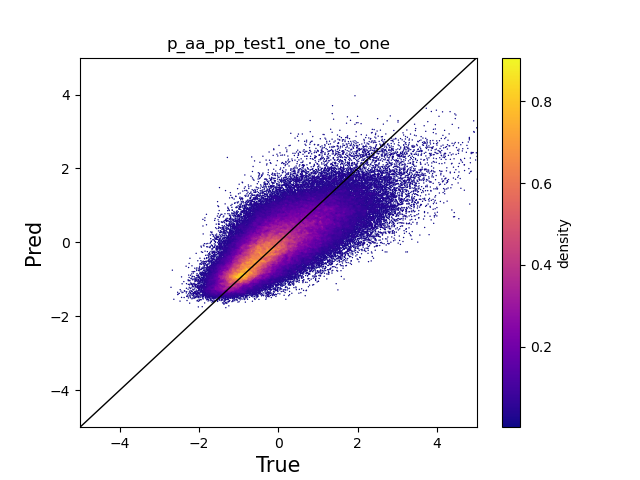}
\includegraphics[width=0.24\textwidth]{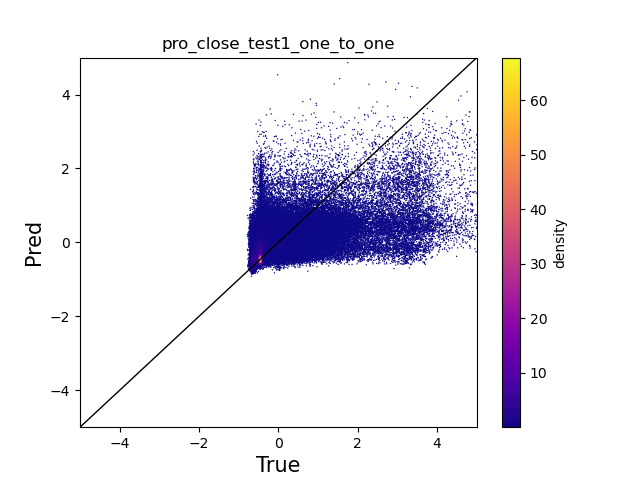}
\includegraphics[width=0.24\textwidth]{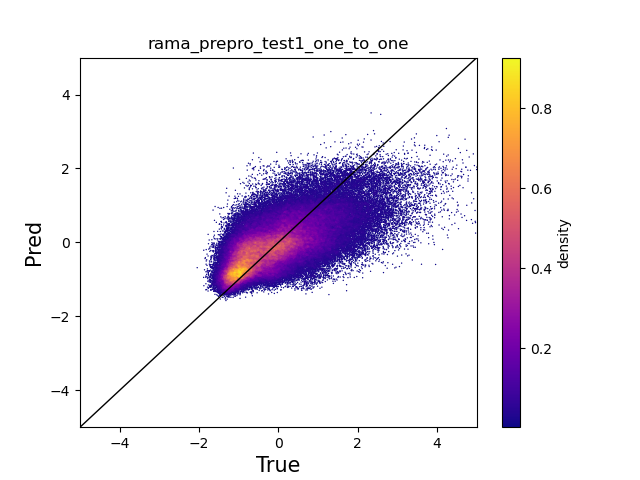}
\includegraphics[width=0.24\textwidth]{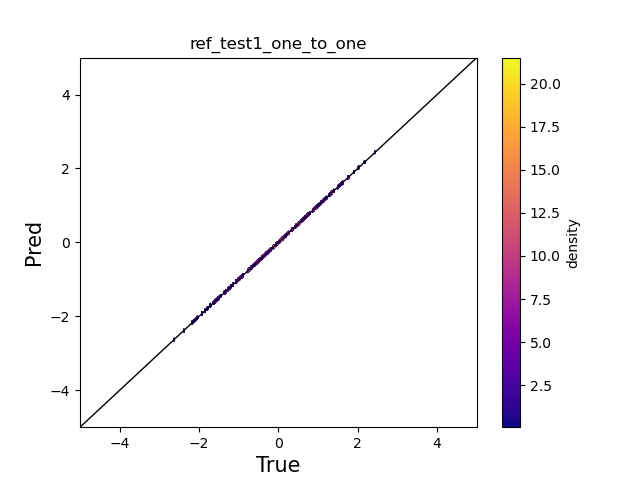}
\includegraphics[width=0.24\textwidth]{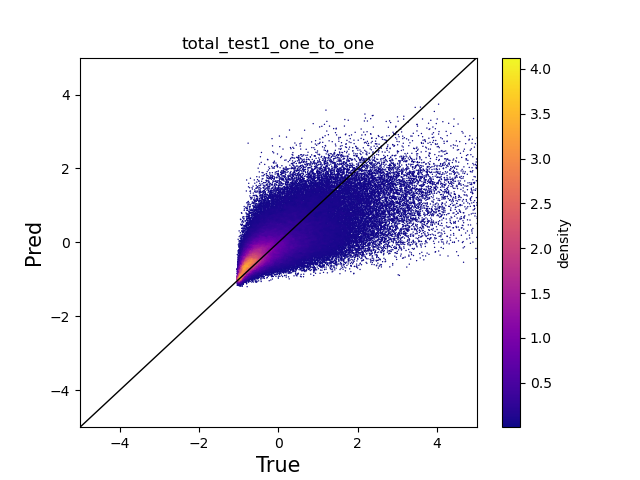}
\end{figure*}

One-to-one plots on the Test1 set are shown in Fig.~\ref{fig:1hhp_one2one}, whose performance is comparable to Test3. Reasonable correlation is observed for most metrics, with exceptional performance on one and lacking performance on a few others.

\begin{figure*}
\caption{Sensitivity analysis plots showing the Pearson correlation coefficients for 1hhp for all predicted scores for Test1 (left), Test2 (center), Test3 (right).}
\includegraphics[width=0.33\textwidth]{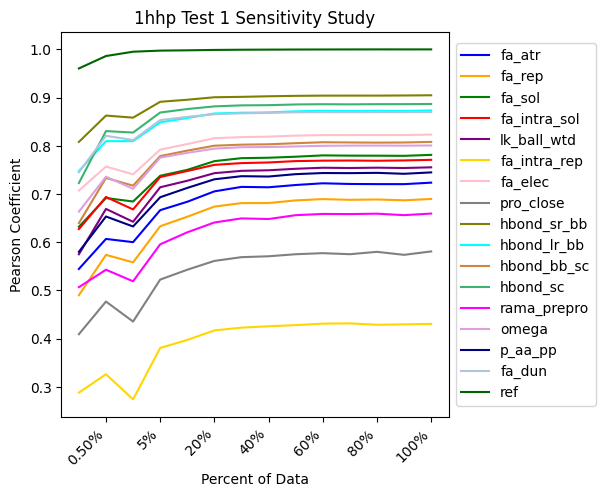}
\includegraphics[width=0.33\textwidth]{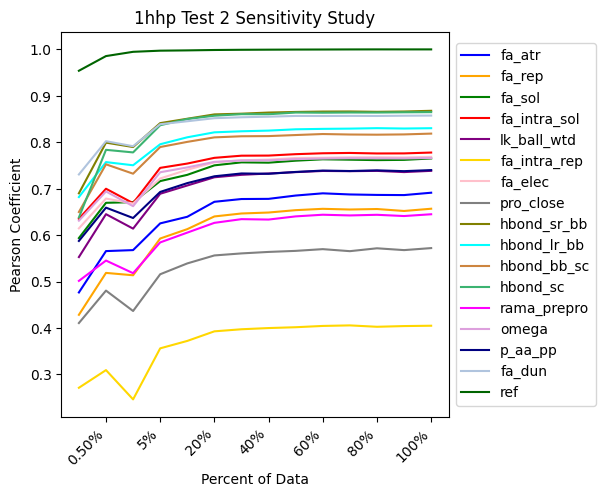}
\includegraphics[width=0.33\textwidth]{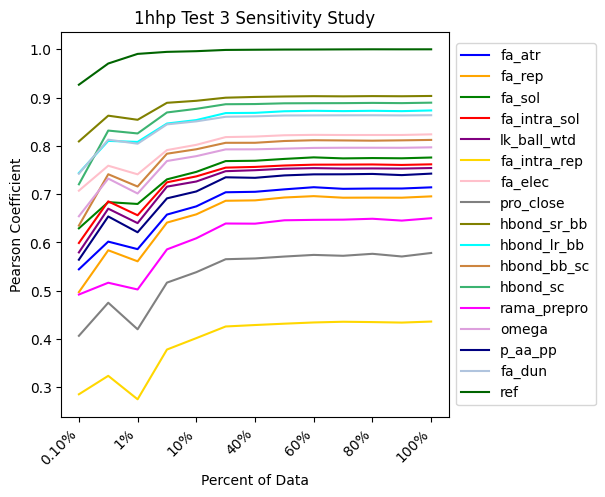}
\label{fig:1hhpline}
\end{figure*}

The sensitivity analysis show that the model trained on only 5\% of the data performs quite similar to performance trained on 100\% of the data as shown in Figure~\ref{fig:1hhpline}. The max difference between any Rosetta score from 100\% and any score from 10\% is $0.159$, the max difference between any Rosetta score from 100\% and any score from 5\% is $0.059$, and finally the maximum difference between any Rosetta score from 100\% and any score from 1\% is $0.156$. The $R^2$ metrics are shown in Table~\ref{tab:1hhpr2}.

\subsection{Results - 1csp}
The second largest protein that we studied is 1csp, the 67-residue structure of \textit{Bacillus subtilis} of \textit{E-coli}. On a single GPU, 1csp required 37 seconds per epoch to train, and trained fully in 5 epochs.

\begin{table}
  \caption{Tests 1, 2 and 3 Pearson correlation coefficients for all predicted scores for 1csp.}
  \label{tab:1csptest}
  \begin{tabular}{llll}
    \toprule
    Rosetta Score/Metric & Test1 & Test2 & Test3 \\
    \midrule
    fa\_atr & 0.742 & 0.721 & 0.734 \\
    fa\_rep & 0.690 & 0.710 & 0.701 \\
    fa\_sol & 0.694 & 0.665 & 0.690 \\
    fa\_intra\_rep  & 0.413 & 0.434 & 0.410 \\
    fa\_intra\_sol   & 0.645 & 0.667 & 0.661 \\
    lk\_ball\_wtd   & 0.473 & 0.409 & 0.470 \\
    fa\_elec    & 0.740 & 0.697 & 0.735 \\
    pro\_close  & 0.601 & 0.656 & 0.608 \\ 
    hbond\_sr\_bb   & 0.606 & 0.644 & 0.603 \\ 
    hbond\_lr\_bb   & 0.872 & 0.843 & 0.872 \\
    hbond\_bb\_sc   & 0.581 & 0.564 & 0.578 \\
    hbond\_sc   & 0.463 & 0.338 & 0.446 \\
    omega   & 0.594 & 0.546 & 0.588 \\
    fa\_dun & 0.847 & 0.870 & 0.847 \\
    p\_aa\_pp   & 0.809 & 0.776 & 0.814 \\
    ref & 1.000 & 1.000 & 0.999 \\
    rama\_prepro    & 0.738 & 0.600 & 0.735 \\
    total   & 0.695 & 0.715 & 0.705 \\
    \bottomrule
  \end{tabular}
\end{table}

\begin{table}
  \caption{Sensitivity analysis showing the $R^2$ metrics for 1csp for Tests 1, 2, and 3.}
  \label{tab:1cspr2}
  \begin{tabular}{llll}
    \toprule
    Percent of Data & Test1 & Test2 & Test3 \\
    \midrule
    100 & 0.476 & 0.442 & 0.468 \\
    90 & 0.475 & 0.445 & 0.467 \\
    80 & 0.476 & 0.441 & 0.468 \\
    70 & 0.474 & 0.442 & 0.465 \\
    60 & 0.472 & 0.439 & 0.462 \\
    50 & 0.469 & 0.435 & 0.459 \\
    40 & 0.464 & 0.429 & 0.456 \\
    30 & 0.464 & 0.426 & 0.454 \\
    20 & 0.448 & 0.409 & 0.438 \\
    10 & 0.429 & 0.387 & 0.417 \\
    5 & 0.404 & 0.363 & 0.390 \\
    1 & 0.321 & 0.292 & 0.313 \\
    0.500 & 0.296 & 0.259 & 0.297 \\
    0.100 & 0.218 & 0.171 & 0.199 \\
    \bottomrule
  \end{tabular}
\end{table}

\begin{figure*}
\caption{One-to-one plots showing the performance on Test1 for 1csp. The x-axis denotes the true values and the y-axis denotes the predictions made by our RoseNet model.} \label{fig:1csp_one2one}
\includegraphics[width=0.24\textwidth]{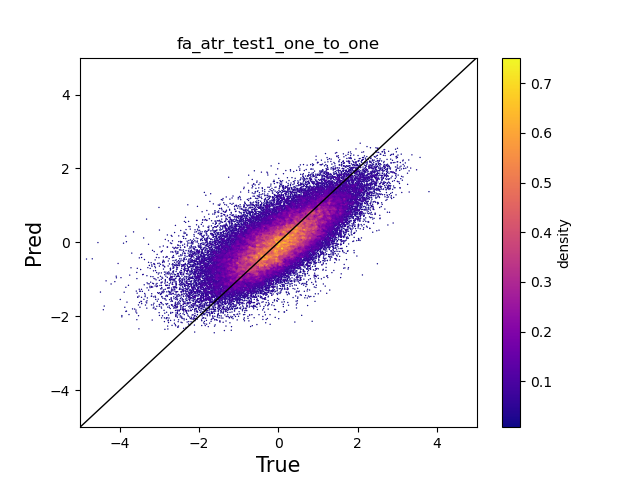}
\includegraphics[width=0.24\textwidth]{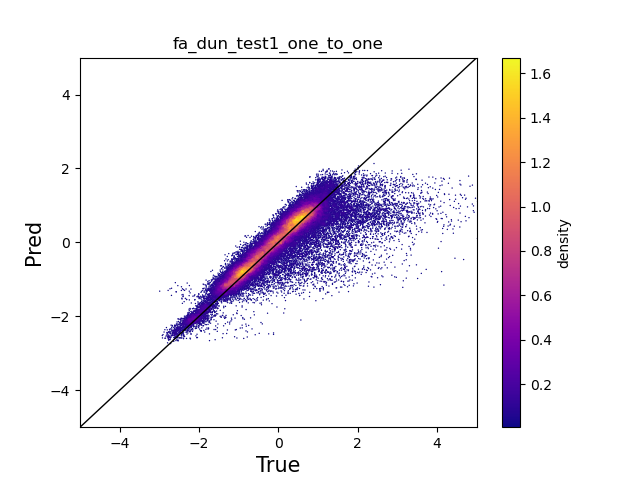}
\includegraphics[width=0.24\textwidth]{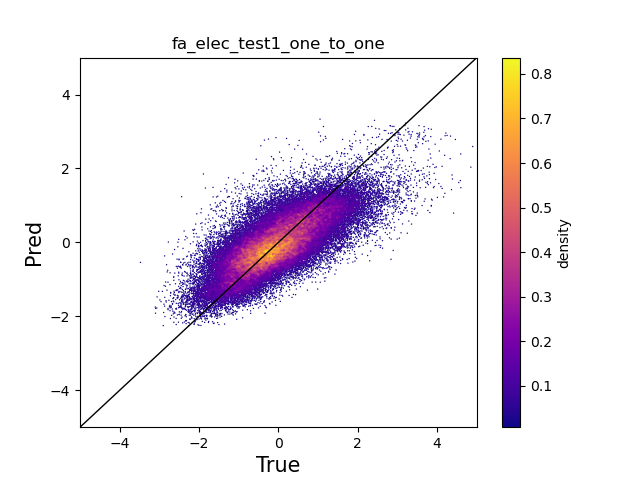}
\includegraphics[width=0.24\textwidth]{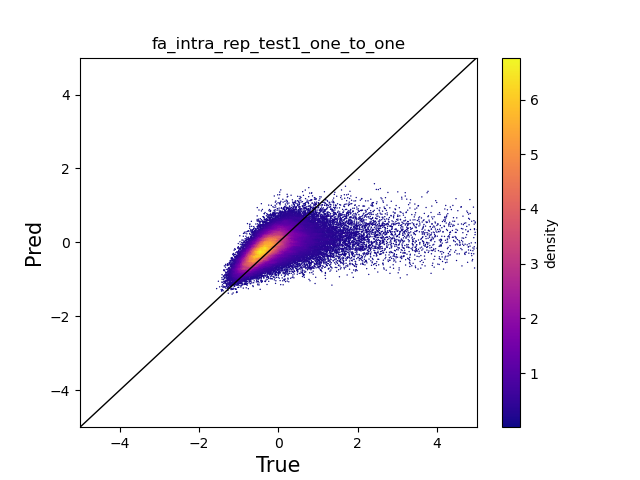}
\includegraphics[width=0.24\textwidth]{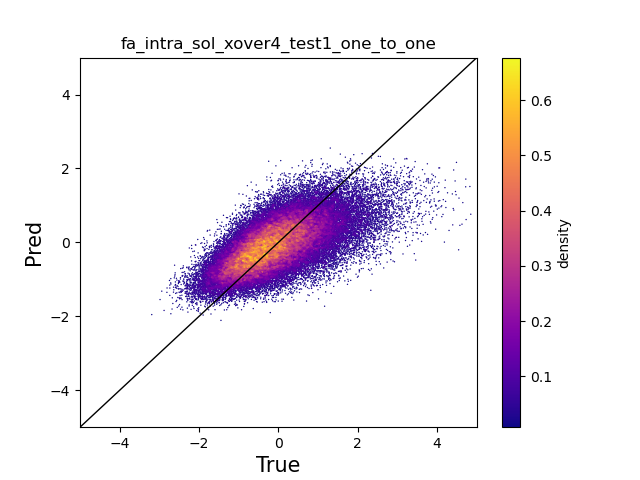}
\includegraphics[width=0.24\textwidth]{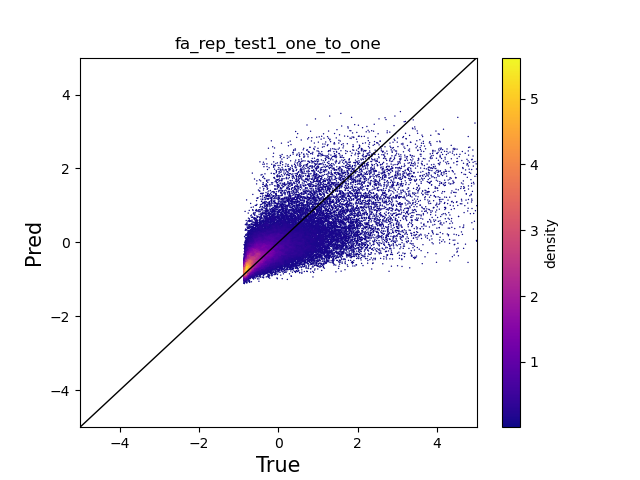}
\includegraphics[width=0.24\textwidth]{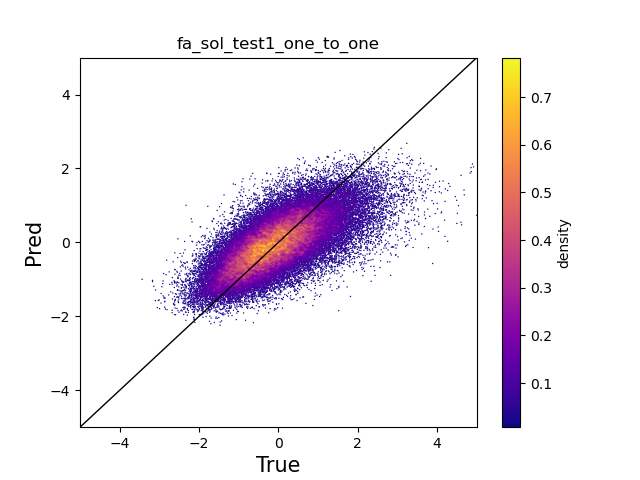}
\includegraphics[width=0.24\textwidth]{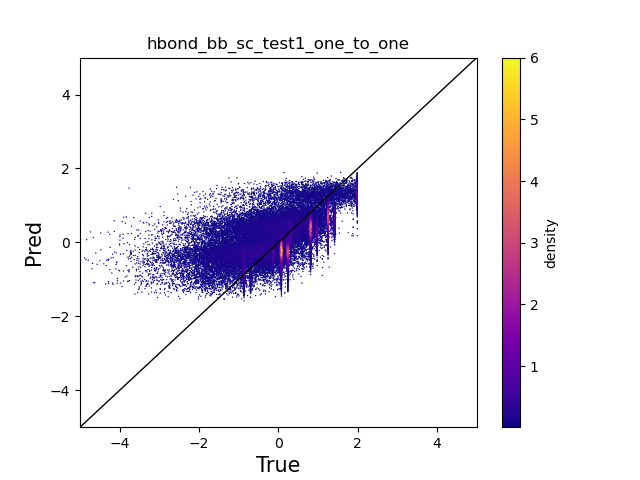}
\includegraphics[width=0.24\textwidth]{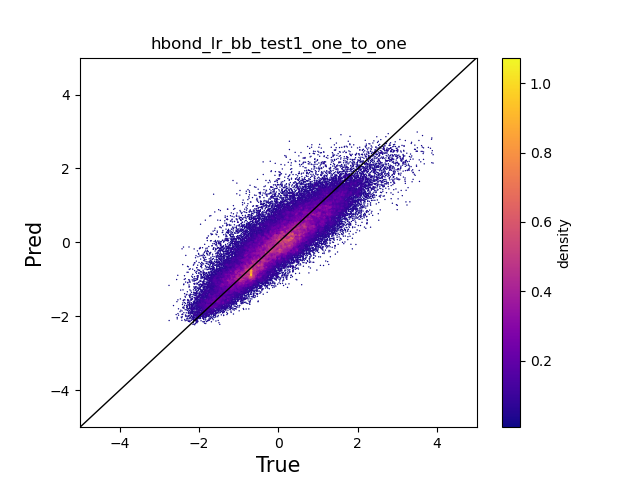}
\includegraphics[width=0.24\textwidth]{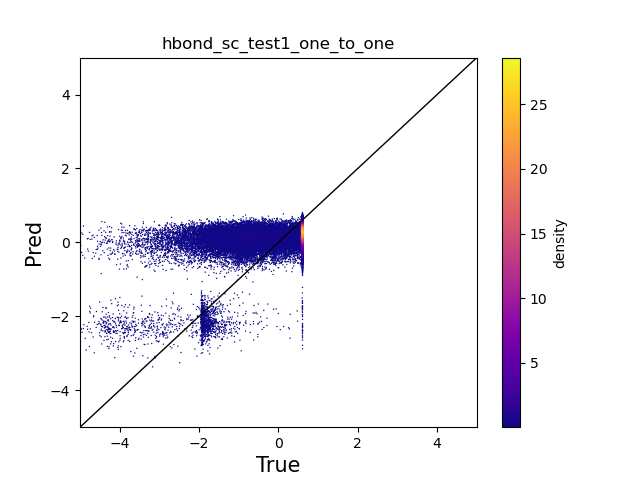}
\includegraphics[width=0.24\textwidth]{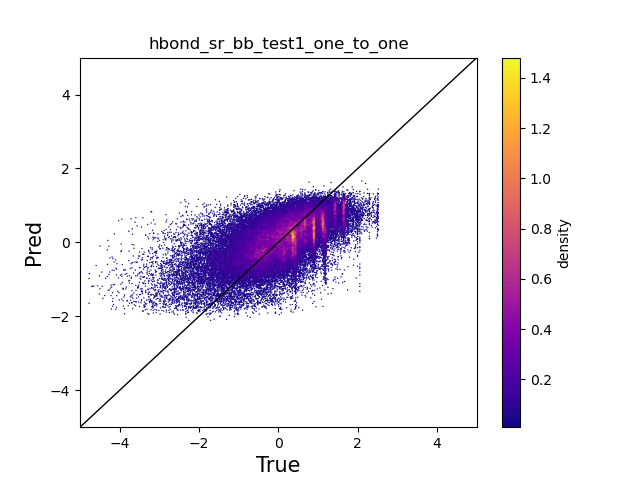}
\includegraphics[width=0.24\textwidth]{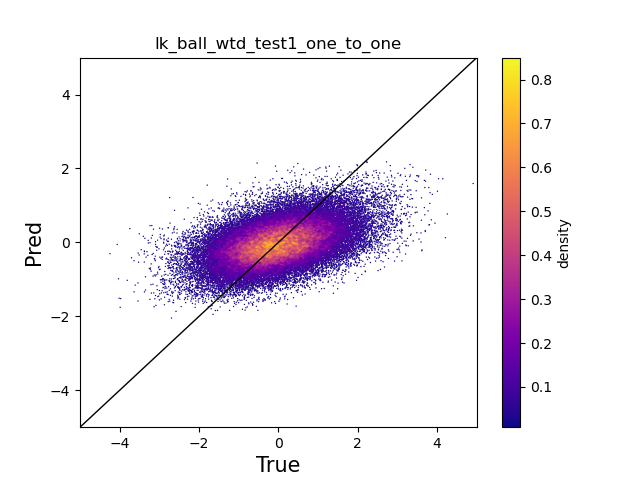}
\includegraphics[width=0.24\textwidth]{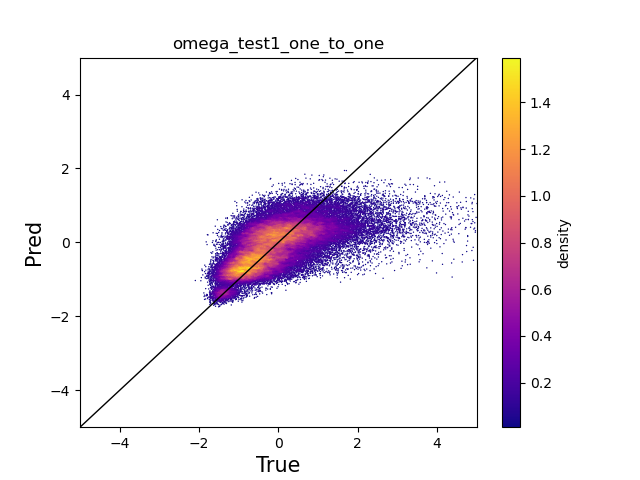}
\includegraphics[width=0.24\textwidth]{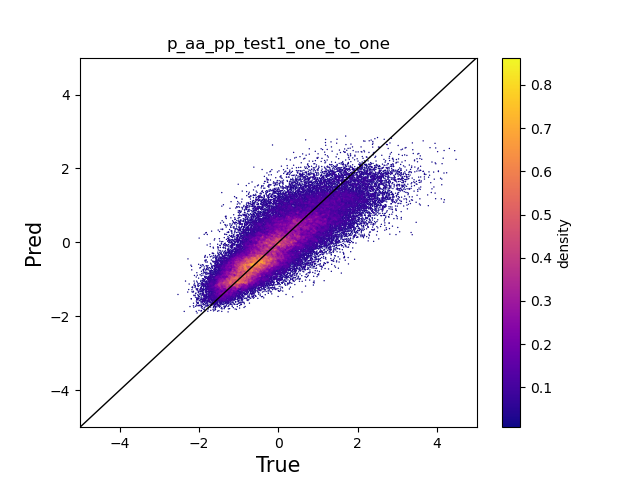}
\includegraphics[width=0.24\textwidth]{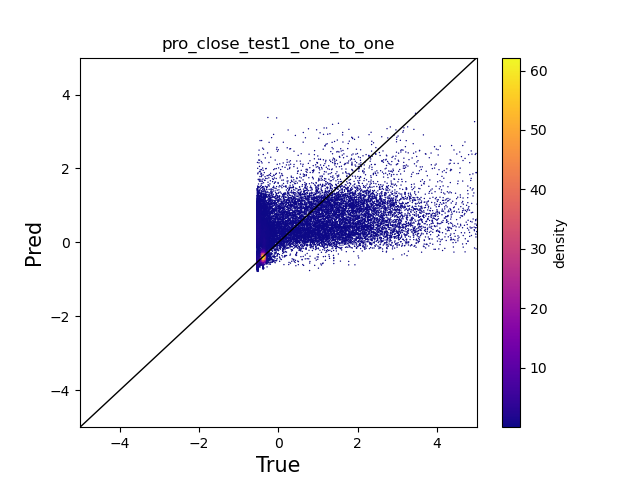}
\includegraphics[width=0.24\textwidth]{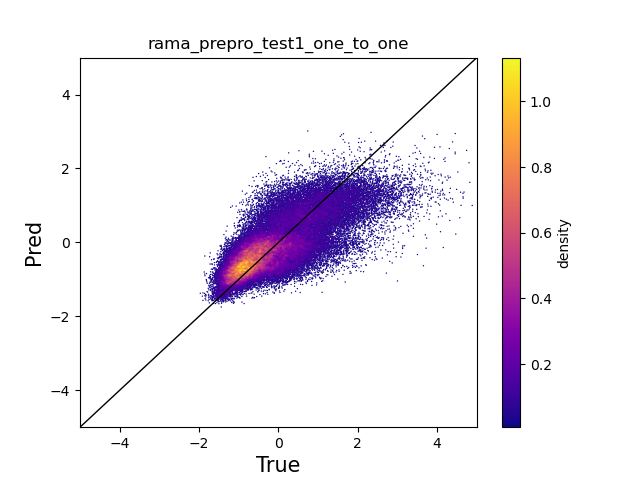}
\includegraphics[width=0.24\textwidth]{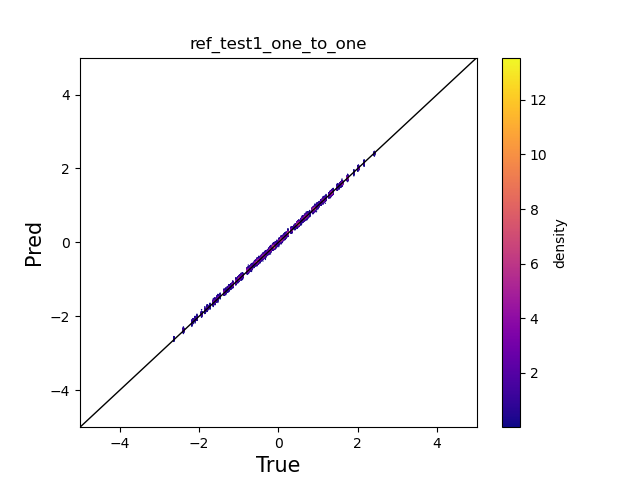}
\includegraphics[width=0.24\textwidth]{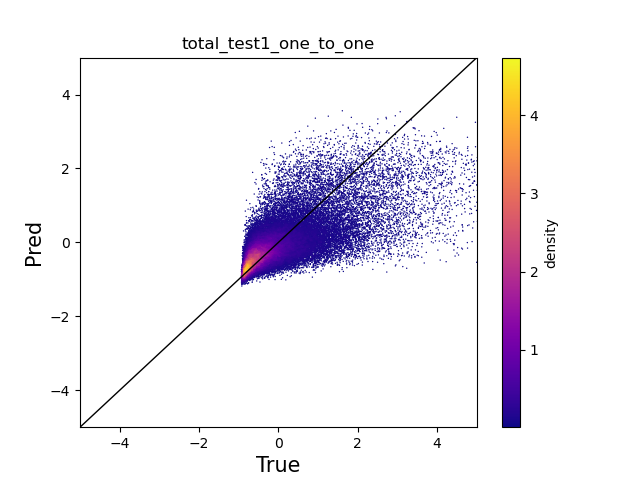}
\end{figure*}

We report results for Test1, Test2 and Test3 in Table~\ref{tab:1csptest}. On the first test set, the scores range from 0.413 for fa\_intra\_rep to 1.000 for the referential energy. The median score is 0.692 and the mean of the scores is 0.678. On the second test set, the scores range from a minimum of 0.338 to a maximum of 1.000. The median score is 0.666 and the mean of all of the scores is 0.658. Compared to 1hhp, this reveals a greater drop-off in performance when generalizing to new insertion-location pairs, but generalization remains quite good.
On the third test set, the scores range from a minimum of 0.410 to a maximum of 0.999. The median score is 0.695 and the mean of all of the scores is 0.678. Here we see no drop-off when generalizing to previously unseen amino-acid pairs, suggesting that the model has learned enough of the amino acids and the behavior in given positions to predict how they will interact. One-to-one plots on the validation set are shown in Fig.~\ref{fig:1csp_one2one}, whose performance is again comparable to Test1. The same general trends are observed.

\begin{figure*}
\caption{Sensitivity analysis plots showing the Pearson correlation coefficients for 1csp for all predicted scores for Test1 (left), Test2 (center), Test3 (right).}
\includegraphics[width=0.33\textwidth]{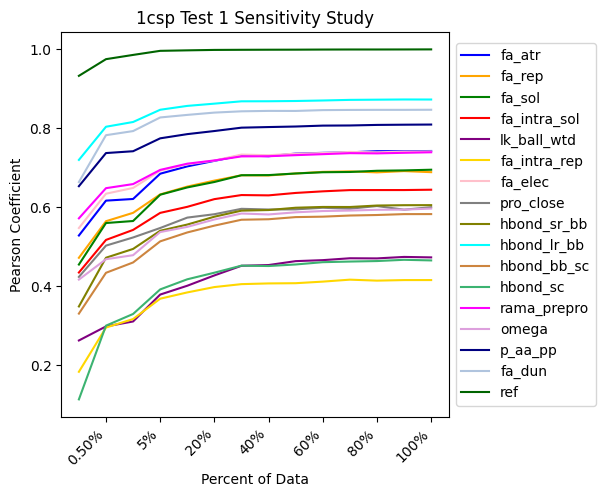}
\includegraphics[width=0.33\textwidth]{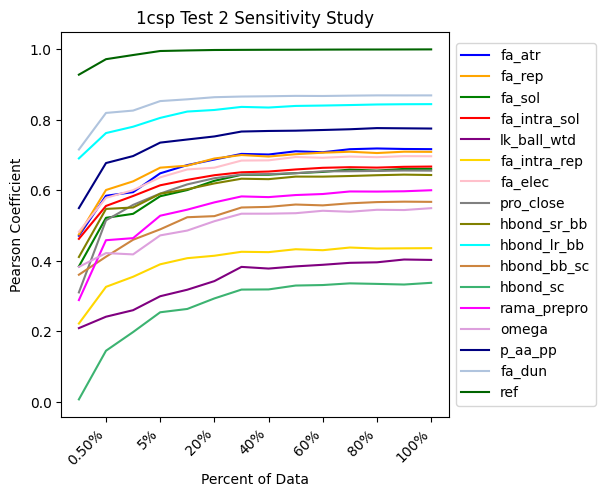}
\includegraphics[width=0.33\textwidth]{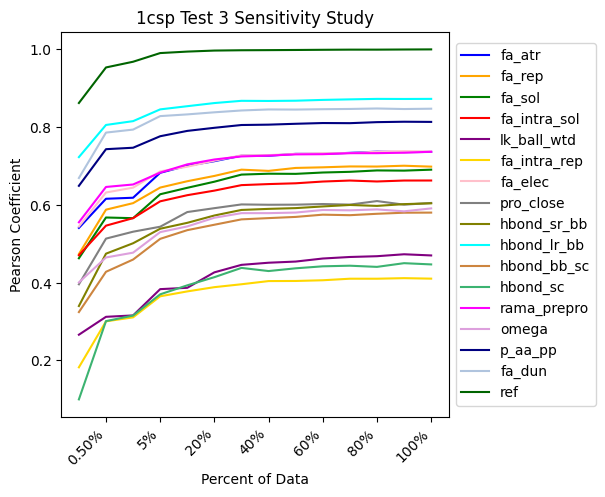}
\label{fig:1cspline}
\end{figure*}

The results of the sensitivity analysis for 1csp show that the model trained on only 10\% of the data performs quite similar to when trained on 100\% of the data (Figure~\ref{fig:1cspline}). The max difference between any Rosetta score from 100\% and any score from 10\% is $0.085$, the max difference between any Rosetta score from 100\% and any score from 5\% is $0.104$, and the maximum difference between any Rosetta score from 100\% and any score from 1\% is $0.160$. The $R^2$ metrics are shown in Table~\ref{tab:1cspr2}.

\subsection{Results - 1crn}
\begin{figure*}
\caption{One-to-one plots showing the performance on Test1 for 1crn. The x-axis denotes the true values and the y-axis denotes the predictions made by our RoseNet model.}
  \label{fig:1crn_one2one}
\includegraphics[width=0.24\textwidth]{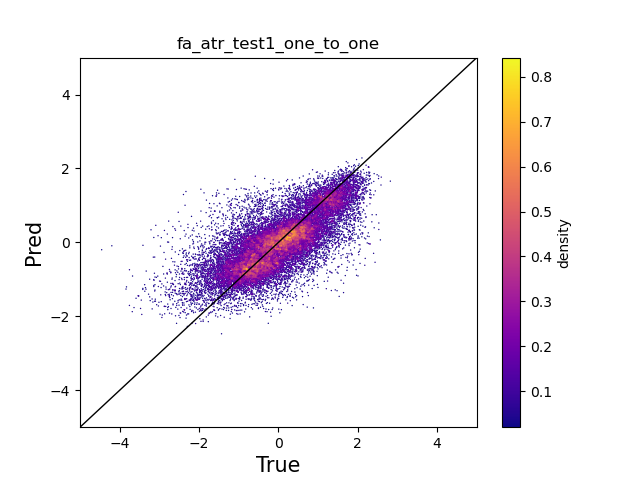}
\includegraphics[width=0.24\textwidth]{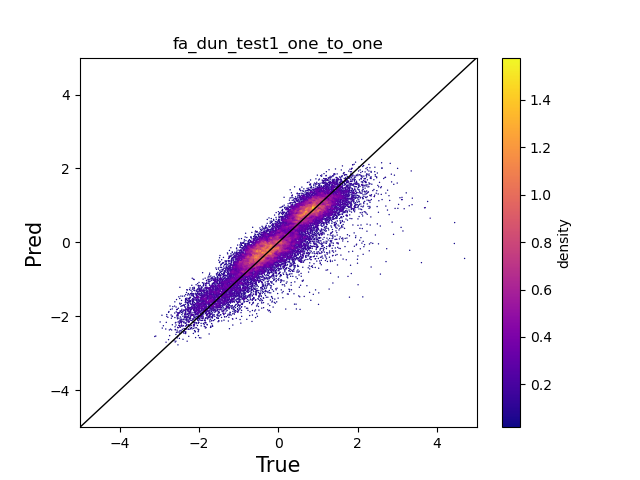}
\includegraphics[width=0.24\textwidth]{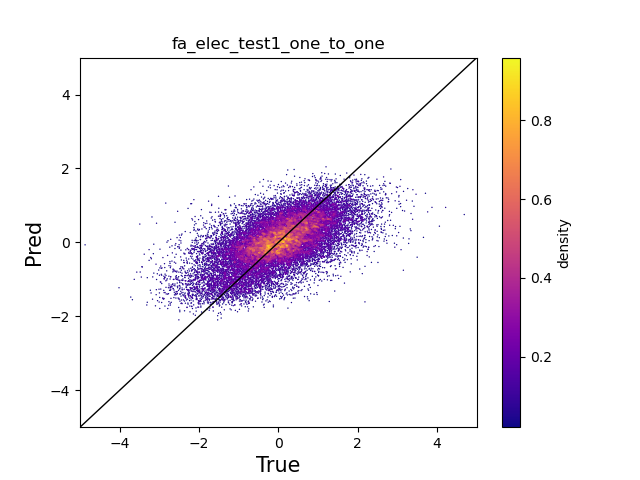}
\includegraphics[width=0.24\textwidth]{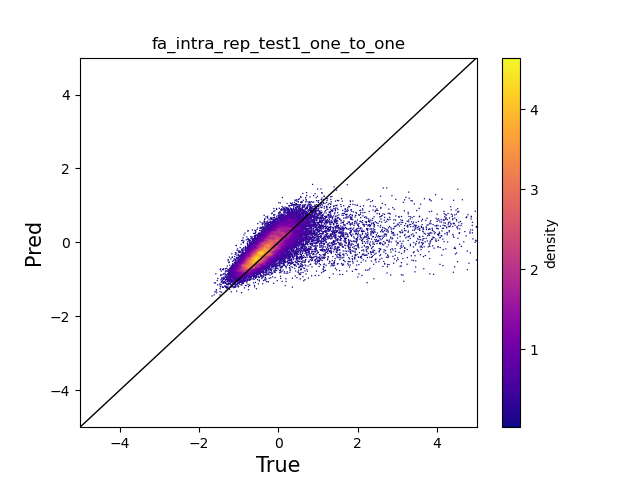}
\includegraphics[width=0.24\textwidth]{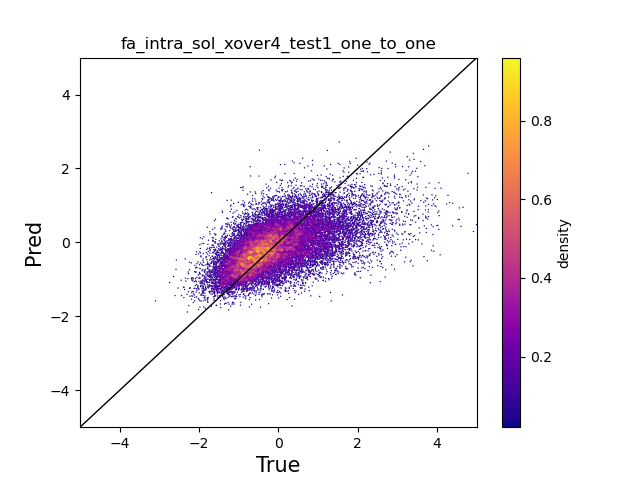}
\includegraphics[width=0.24\textwidth]{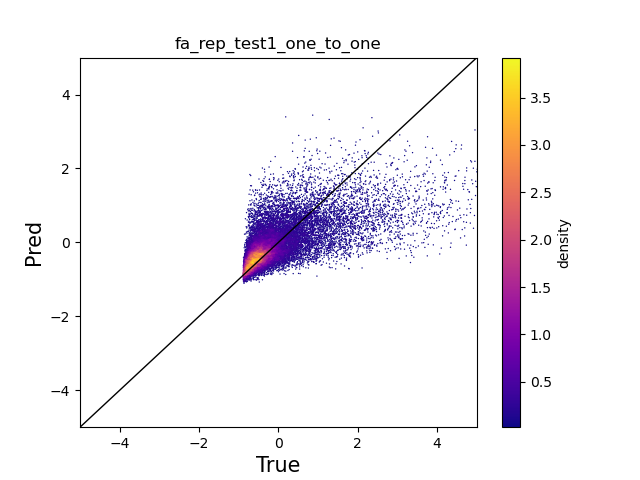}
\includegraphics[width=0.24\textwidth]{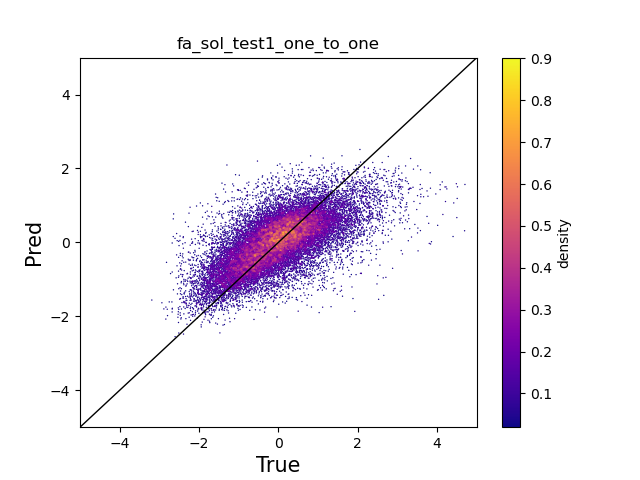}
\includegraphics[width=0.24\textwidth]{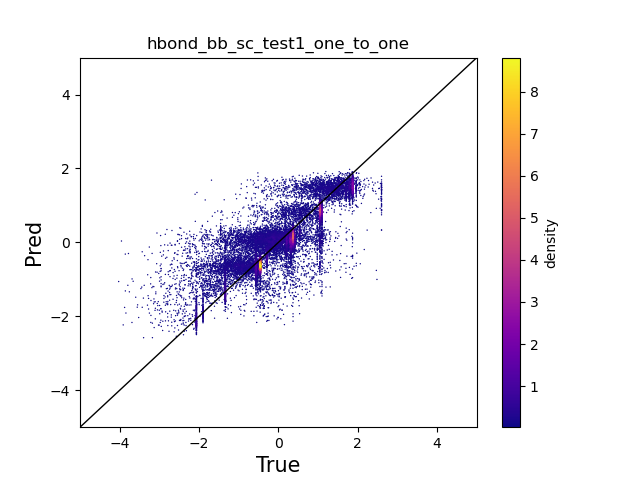}
\includegraphics[width=0.24\textwidth]{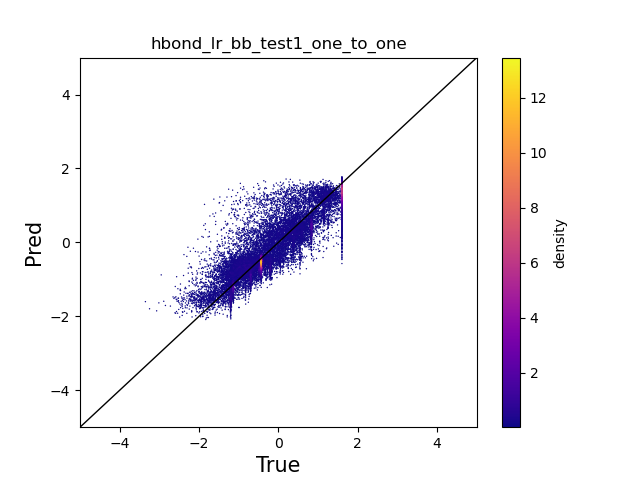}
\includegraphics[width=0.24\textwidth]{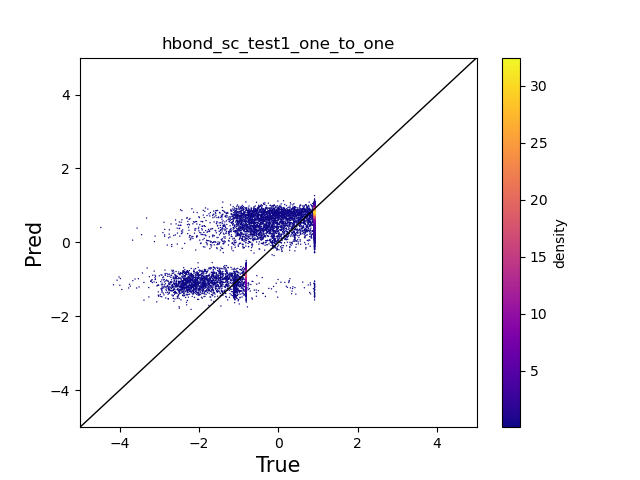}
\includegraphics[width=0.24\textwidth]{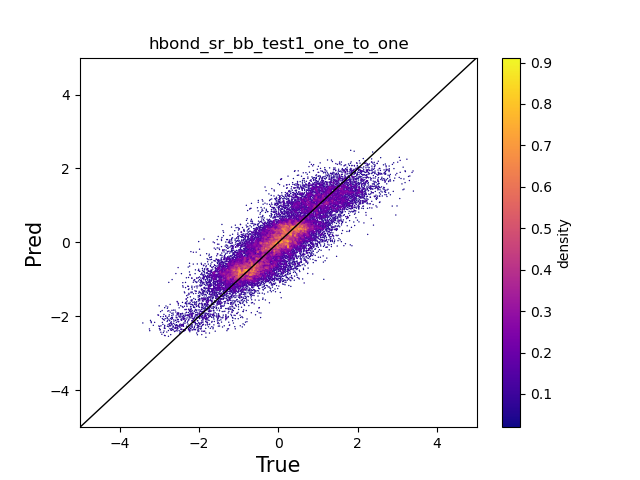}
\includegraphics[width=0.24\textwidth]{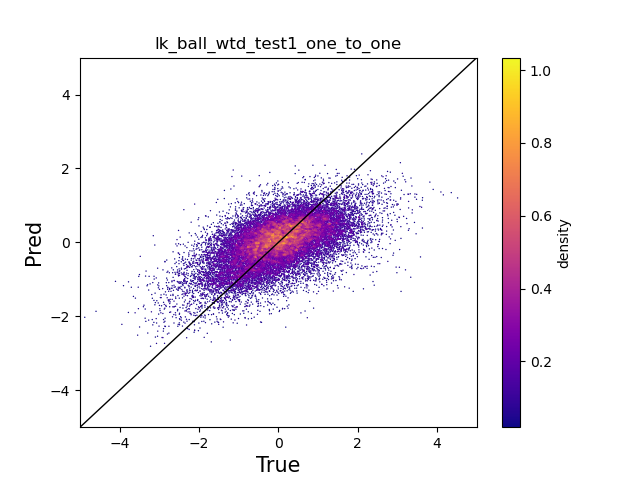}
\includegraphics[width=0.24\textwidth]{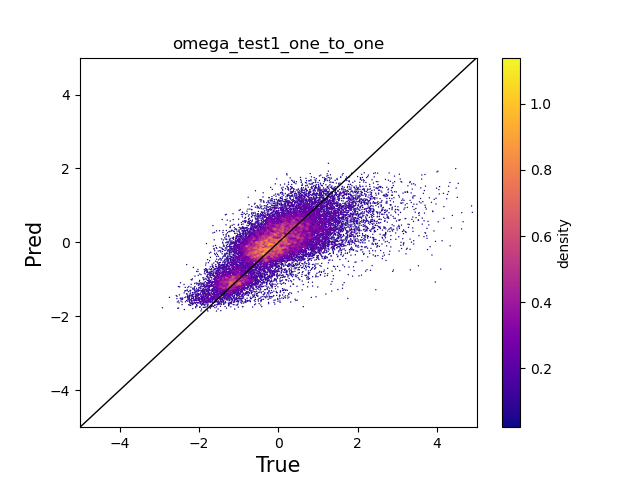}
\includegraphics[width=0.24\textwidth]{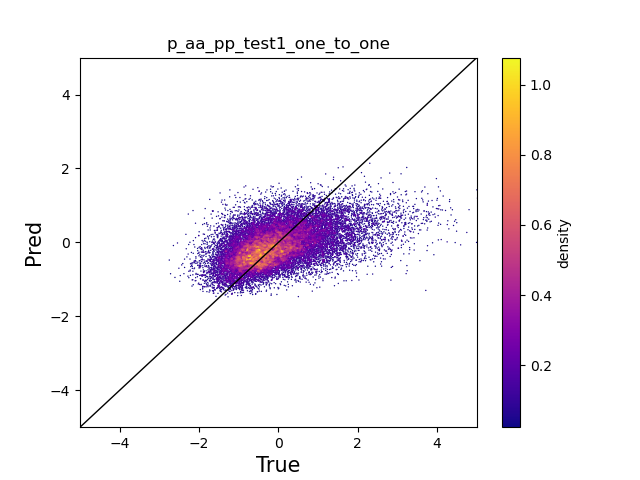}
\includegraphics[width=0.24\textwidth]{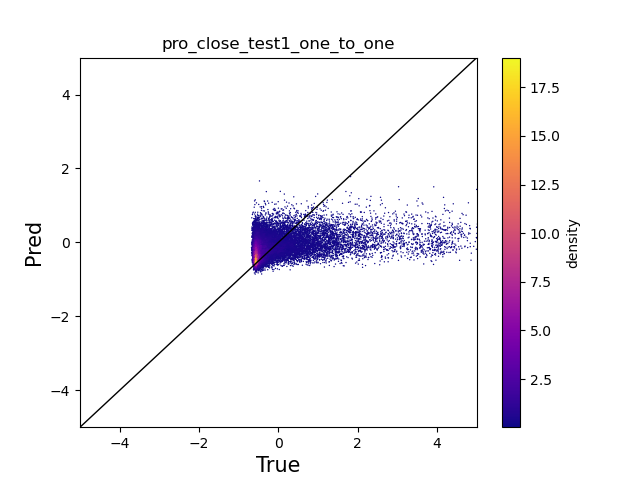}
\includegraphics[width=0.24\textwidth]{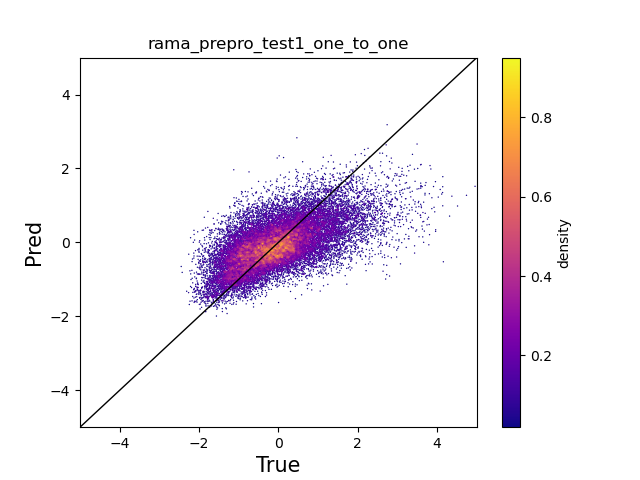}
\includegraphics[width=0.24\textwidth]{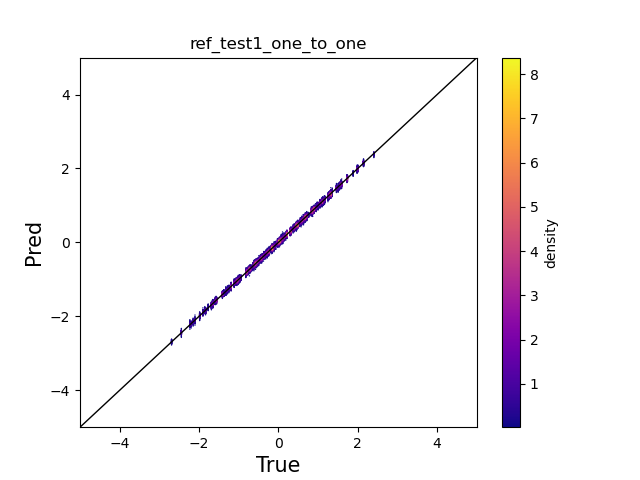}
\includegraphics[width=0.24\textwidth]{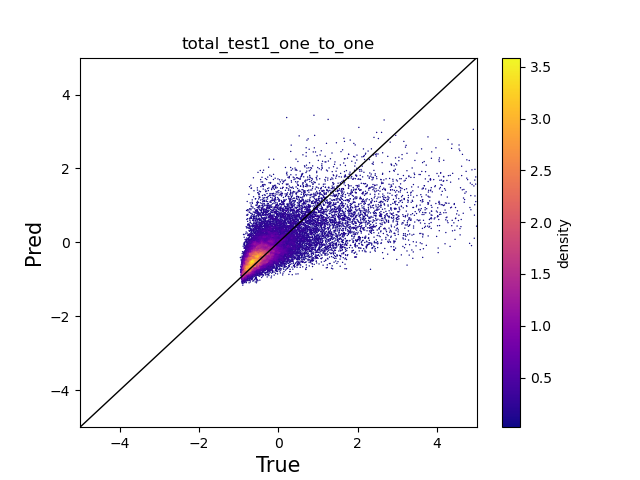}
\end{figure*}

The smallest protein we studied, PDB structure 1crn, is the 46-amino acid (48 after the double insertions) structure of Crambin. On a single GPU, 1crn takes 14 seconds per epoch to train, and trains fully in 4 epochs. Test set results are shown in Table~\ref{tab:1crntest}, and one-to-one plots on the validation set are shown in Fig.~\ref{fig:1crn_one2one}.

\begin{table}
  \caption{Tests 1, 2 and 3 Pearson correlation coefficients for all predicted scores for 1crn.}
  \label{tab:1crntest}
  \begin{tabular}{llll}
    \toprule
    Rosetta Score/Metric & Test1 & Test2 & Test3 \\
    \midrule
    fa\_atr & 0.733 & 0.597 & 0.734 \\
    fa\_rep & 0.641 & 0.608 & 0.638 \\
    fa\_sol & 0.655 & 0.580 & 0.652 \\
    fa\_intra\_rep & 0.508 & 0.479 & 0.540 \\
    fa\_intra\_sol & 0.572 & 0.576 & 0.548 \\
    lk\_ball\_wtd & 0.575 & 0.536 & 0.599 \\
    fa\_elec & 0.607 & 0.452 & 0.613 \\
    pro\_close & 0.382 & 0.384 & 0.390 \\
    hbond\_sr\_bb & 0.873 & 0.814 & 0.871 \\
    hbond\_lr\_bb & 0.884 & 0.760 & 0.876 \\
    hbond\_bb\_sc & 0.807 & 0.765 & 0.803 \\
    hbond\_sc & 0.820 & 0.752 & 0.817 \\
    omega & 0.711 & 0.600 & 0.706 \\
    fa\_dun & 0.891 & 0.892 & 0.895 \\
    p\_aa\_pp & 0.488 & 0.505 & 0.501 \\
    ref & 0.999 & 0.999 & 0.999 \\
    rama\_prepro & 0.617 & 0.588 & 0.635 \\
    total & 0.649 & 0.610 & 0.647 \\
    \bottomrule
  \end{tabular}
\end{table}

\begin{table}
  \caption{Sensitivity analysis showing the $R^2$ metrics for 1crn for Tests 1, 2, and 3.}
  \label{tab:1crnr2}
  \begin{tabular}{llll}
    \toprule
    Percent of Data & Test1 & Test2 & Test3 \\
    \midrule
    100 & 0.494 & 0.382 & 0.484 \\
    90 & 0.497 & 0.388 & 0.485 \\
    80 & 0.495 & 0.383 & 0.483 \\
    70 & 0.493 & 0.383 & 0.473 \\
    60 & 0.469 & 0.353 & 0.457 \\
    50 & 0.471 & 0.350 & 0.456 \\
    40 & 0.479 & 0.362 & 0.464 \\
    30 & 0.477 & 0.361 & 0.466 \\
    20 & 0.457 & 0.348 & 0.445 \\
    10 & 0.416 & 0.297 & 0.407 \\
    5 & 0.390 & 0.275 & 0.372 \\
    1 & 0.332 & 0.227 & 0.317 \\
    0.500 & 0.320 & 0.207 & 0.302 \\
    0.100 & 0.095 & 0.049 & 0.090 \\
    \bottomrule
  \end{tabular}
\end{table}

On the first test set, the scores range from 0.382 for pro\_close to 0.999 for the referential energy. The median score is 0.652 and the mean of the scores is 0.690. 
On the second test set, the scores range from a minimum of 0.384 to a maximum of 0.999. The median score is 0.599 and the mean of all of the scores is 0.639. As with 1csp, we see some drop in performance when generalizing to these unseen insertion-location pairs, but it remains good.
On the third test set, the scores range from a minimum of 0.390 to a maximum of 0.999. The median score is 0.650 and the mean of all of the scores is 0.693. Again following the trend of 1csp, we see essentially no drop-off in performance from Test1 to Test3, suggesting that it is relatively easy to generalize to unseen amino-acid pairs. 

\begin{figure*}
\caption{Sensitivity analysis plots showing the Pearson correlation coefficients for 1crn for all predicted scores for Test1 (left), Test2 (center), Test3 (right).}
\includegraphics[width=0.33\textwidth]{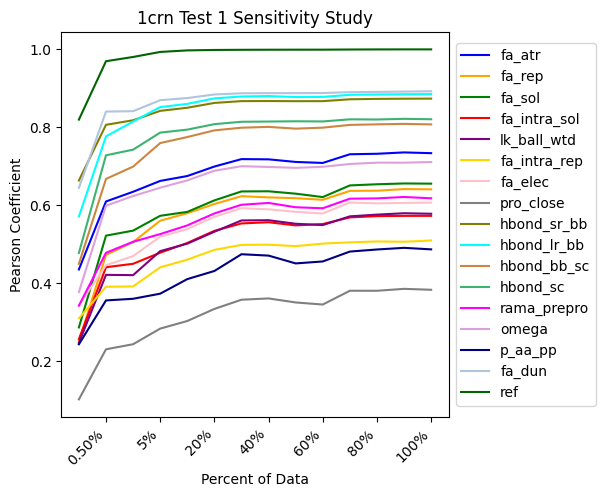}
\includegraphics[width=0.33\textwidth]{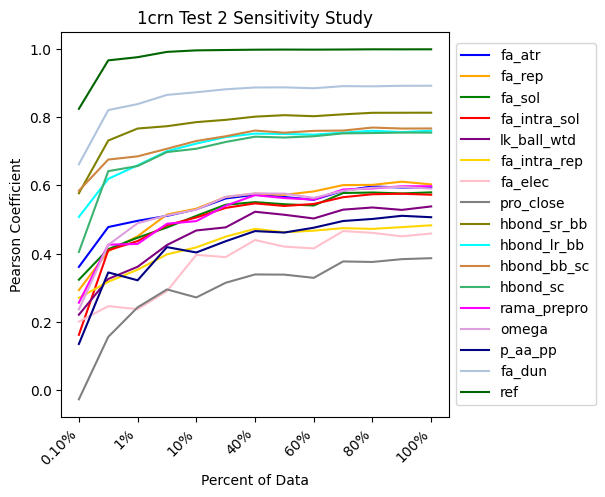}
\includegraphics[width=0.33\textwidth]{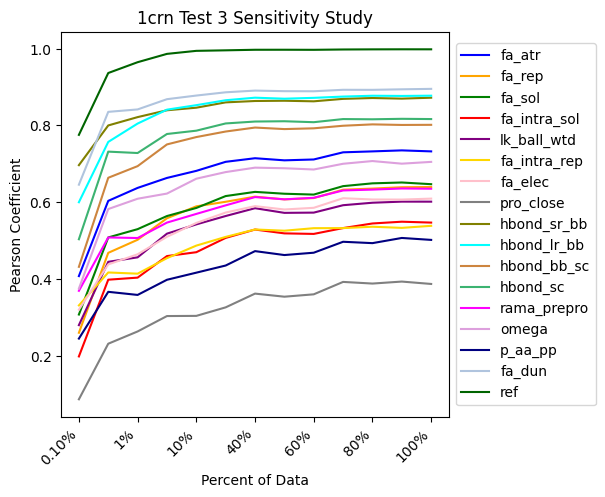}
\label{fig:1crnline}
\end{figure*}

The results of the sensitivity analysis for 1crn show that the model trained on only 10\% of the data performs quite similar to performance trained on 100\% of the data as shown in Figure~\ref{fig:1crnline}. 
The max difference between any Rosetta score from 100\% and any score from 10\% is $0.115$, the max difference between any Rosetta score from 100\% and any score from 5\% is $0.161$, and the maximum difference between any Rosetta score from 100\% and any score from 1\% is $0.223$. The $R^2$ metrics are shown in Table~\ref{tab:1crnr2}.

\section{Discussion}
For each protein, a model was trained with varied depth (up to 10 RoseNet blocks), and in each case, the best performing model contained only one RoseNet block. 
Other hyperparameters were varied as well, such as the optimizer, for which the best performance was seen with NAdam, and the learning rate, for which the model performance was best at 0.0001 across the three proteins. 

As was noted in the previous section, we typically see a drop in performance on the second test data set, when compared with the first and third sets. This implies that the model was more readily able to generalize previously unseen amino acid residues than to previously unseen insertion positions. There were, however, isolated cases of metrics for certain proteins where the Test2 predictions outperformed the Test3 and Test1 predictions, such as with  fa\_intra\_rep, fa\_intra\_sol, pro\_close, hbond\_sr\_bb, fa\_dun and total for the 1csp protein.

As expected, the model trained on the largest protein typically predicted more accurately for its respective metrics. However, there were a few metrics for which this wasn't the case. fa\_intra\_rep, the Lennard-Jones repulsive between atoms in the same residue, was the most poorly performing metric for the two larger proteins. However for 1crn (the smallest protein), this metric substantially improved on the 1csp and 1hhp results across all test sets in the Pearson correlation coefficient. In the third test set specifically, for this metric, the performance jumped from 0.410 in 1csp and 0.437 in 1hhp to as high as 0.540 for the smallest protein, 1crn. The reason why for some metrics the smaller proteins performed better than for 1hhp remains unknown, but may be caused by the different count and distributions of secondary structures across the studied proteins. For example 1csp is the only protein of the three without an $\alpha$-helix. Similarly, more than half of 1crn's residues are members of an $\alpha$-helix, while fewer than 10\% of the residues in 1hhp and 1csp are in $\alpha$-helices.

The metric that performed well most consistently (excepting ref) across all proteins and test sets was fa\_dun, which measures the internal energy of sidechains. hbond\_lr\_bb also performed well.

\section{Future Work}
In future work, one could use a transformer neural network to predict Rosetta metrics for unseen mutant sequences with two InDels. The entire amino acid sequence might be utilized as the input to the transformer as these sequences hold valuable information about protein structure and function. Transformers excel in capturing distant dependencies and patterns within sequences, making them well-suited for analyzing the complete amino acid chain. These results could also be replicated on additional proteins.

We also leave for future work a systematic investigation to determine the extent to which variations in secondary structure or domain features across proteins contribute to variations in accuracy of predictions.

\bibliographystyle{ACM-Reference-Format}
\bibliography{references}

\end{document}